\documentclass[aps,physrev,preprint,superscriptaddress]{revtex4-2}
\usepackage{graphicx}

\usepackage{soul} %highlight
\usepackage{xcolor} %highlight
\sethlcolor{yellow} %highlight

\begin{document}

%\preprint{}

\title{Predicting Interface Structure using the Minima Hopping Method with a Machine Learning Interatomic Potential}


\author{Chang-Ti Chou}
\affiliation{%
  Department of Materials Science and Engineering, Northwestern University, Evanston, IL 60208, United States}%
  
\author{Menghang Wang}
\affiliation{%
  Harvard John A. Paulson School of Engineering and Applied Sciences,
  Harvard University, Cambridge, MA 02138, USA}%

\author{Chao Yang}
\affiliation{%
  Max Planck Institute for Solid State Research, Stuttgart 70569, Germany}%

\author{Peter A. van Aken}
\affiliation{%
  Max Planck Institute for Solid State Research, Stuttgart 70569, Germany}%

\author{Nicola H. Perry}%
\affiliation{%
  Department of Materials Science and Engineering, The Grainger College of Engineering, University of Illinois Urbana-Champaign, Urbana, IL, USA}%
\affiliation{%
  Materials Research Laboratory, The Grainger College of Engineering, University of Illinois Urbana-Champaign, Urbana, IL, USA}%

\author{Boris Kozinsky}
\affiliation{%
  Harvard John A. Paulson School of Engineering and Applied Sciences,
  Harvard University, Cambridge, MA 02138, USA}
\affiliation{%
  Robert Bosch Research and Technology Center, Watertown, MA 02472, USA}
  
\author{Christopher M. Wolverton}%
\email{Contact author: c-wolverton@northwestern.edu}
\affiliation{%
  Department of Materials Science and Engineering, Northwestern University, Evanston, IL 60208, United States}%

%\thanks{}


\date{\today}

\begin{abstract}

Predicting atomic-scale interfacial structures remains a central challenge in materials science due to their structural complexity and the difficulty of direct comparison between computational and experimental results. In this study, we present an efficient approach for interface structure prediction that integrates the Minima Hopping Method (MHM) with the state-of-the-art machine learning interatomic potential (MLIP), Allegro. We demonstrate that the MHM–Allegro approach provides a robust and computationally efficient approach for predicting interfacial structures in the representative benchmark system, SrTiO$_3$ $\Sigma3$(112)[110] tilt grain boundaries (GBs), consistently identifying the lowest-energy configurations across different stoichiometries. Furthermore, we introduce a novel strategy for constructing defect-representative training datasets without explicitly including defective configurations, achieving excellent extrapolative performance in interface predictions. The predictive capability of the approach is further validated through direct comparison with experimental observations of the SrTiO$_3$ $\Sigma5$(310)[001] GB, where the predicted atomic configurations show strong agreement with experimental measurements. This work represents a significant step toward bridging the gap between \textit{ab initio} predictions and experimentally observed interfacial structures.

\end{abstract}

%\keywords{}

\maketitle

\section{\label{sec:intro}Introduction}

Solid-solid interfaces, which are three-dimensional defects separating distinct crystallographic regions or phases, play a pivotal role in governing material behavior, exerting profound influence over mechanical \cite{hirth1972influence,watanabe1999control}, electronic \cite{tapaszto2012mapping,yang2021determination}, optical \cite{hunderi1973influence,kazmerski1983optical}, and magnetic \cite{bonetti1999disordered,sasaki2016structure} properties. Among the various types of interfaces, grain boundaries \cite{priester2012grain, rohrer2011grain, watanabe2011grain, quirk2024grain} (GBs) are among the most extensively studied and lie at the intersection of two misoriented grains of the same crystalline phase. The disordered bonding environments and symmetry breaking at interfaces give rise to unique thermodynamic and kinetic behaviors that differ markedly from those of the bulk. These interfacial regions often constitute the kinetic bottleneck in energy-relevant processes \cite{defferriere2022photo}, such as lithium-ion transport in solid-state batteries \cite{he2021grain,symington2021elucidating,yan2018tailoring} and proton diffusion in fuel cell materials \cite{shirpour2011nonlinear,bi2018tailoring,obewhere2024engineering}. Consequently, elucidating the atomic-scale structure of interfaces is essential for advancing both the fundamental understanding of materials and the design of next-generation energy technologies.

%difficulty for predict interface computaionally, and hard to compare to exp
Despite their importance, the prediction of atomic-scale interfacial and grain boundary structures remains far less developed than that of bulk crystal structures. This disparity primarily arises from the high computational cost and configurational complexity of interface modeling. Accurate simulations typically require large supercells containing hundreds of atoms in order to minimize artifacts from periodic boundary conditions. Moreover, structurally each grain boundary exhibits five degrees of freedom \cite{quirk2024grain}, resulting in an enormous configurational space to explore even within a single material system. These factors render exhaustive high-throughput \textit{ab initio} screening infeasible. Even when prediction can be achieved, direct comparison between theoretical structures and arbitrary experimentally observed interfaces remains challenging. Computational predictions typically correspond to 0~K equilibrium configurations, whereas experimentally observed interfaces reflect finite-temperature effects as well as kinetic and entropic contributions introduced during synthesis and subsequent cooling. These factors complicate direct comparisons between theory and experiment. Consequently, well-controlled bicrystal samples provide a more suitable benchmark for comparing experimental observations with theoretical predictions. By fixing the relative grain orientation and chemical composition, and applying sufficiently long annealing treatments, such samples can approach near-equilibrium structures, enabling more meaningful comparison with 0 K theoretical predictions.

To mitigate the computational expense of \textit{ab initio} methods, early studies employed empirical interatomic potentials to approximate the potential energy surface (PES). These approaches proved successful primarily for single-component systems. For example, von Alfthan \textit{et al.} investigated ordering phenomena at $\Sigma5$(001) twist grain boundaries in silicon using the Tersoff potential \cite{von2006structures}, while Tschopp \textit{et al.} explored asymmetric tilt grain boundaries in copper and aluminum by varying atomic density and cell shifts near the boundary \cite{tschopp2007asymmetric,tschopp2007structures}. Recent efforts have increasingly focused on complex multi-component oxide systems such as TiO$_2$ \cite{gao2019interface,schusteritsch2021anataselike} and SrTiO$_3$ \cite{chua2010genetic,schusteritsch2014predicting,zhao2014interface,chou2025predicting}. Among these, SrTiO$_3$ grain boundaries have emerged as benchmark systems for evaluating interface prediction methodologies in multicomponent materials. In these systems, the presence of multiple atomic species substantially enlarges the configurational search space and necessitates consideration of non-stoichiometric interfaces, where the local composition at the boundary departs from the bulk stoichiometry. Such non-stoichiometric interfaces have been observed experimentally \cite{yang2021determination}, and computational studies have demonstrated that they can be thermodynamically more stable than their stoichiometric counterparts under certain chemical potentials \cite{yang2013quantifying,kim2015study}. While density functional theory (DFT) often provides an accurate method for describing these oxide systems, its prohibitive computational cost limits direct use in large-scale interface searches. Consequently, hybrid approaches have emerged, combining generative structure-search algorithms with surrogate PES models to identify low-energy configurations efficiently. After structural exploration, candidate configurations are typically refined through DFT relaxation to obtain accurate interfacial energies. Among these hybrid approaches, the Minima Hopping Method \cite{chou2025predicting} (MHM) has proven to be a robust interface searching algorithm. The accuracy and efficiency of such approaches depend critically on the quality of the surrogate PES model. For instance, the adaptive genetic algorithm (AGA) proposed by Zhao \textit{et al.} \cite{zhao2014interface} enhanced exploration performance by iteratively updating the embedded atom method (EAM) potential, outperforming traditional genetic algorithms (GA) relying solely on fixed Buckingham potential \cite{chua2010genetic}. 

Over the past decade, machine learning (ML) has emerged as a transformative tool across a broad range of scientific disciplines, including materials science. In the atomic scale interface prediction, two main ML-based paradigms have emerged: (1) generative models, such as diffusion-based frameworks \cite{lei2024grand}, and (2) machine learning interatomic potentials (MLIPs) serving as surrogate PES models \cite{batzner20223,musaelian2023learning,chen2022universal,deng2023chgnet} in the hybrid approaches. While generative models hold promise for direct structure generation, their practical utility remains limited by the scarcity of comprehensive datasets of interfacial configurations. In contrast, MLIPs, when coupled with structure-search algorithms such as MHM, enable efficient and accurate exploration of interfacial energy landscapes with near-DFT fidelity. The recent development of universal MLIPs (uMLIPs) \cite{chen2022universal,deng2023chgnet,kaplan2025foundational}, trained on large-scale databases such as the Open Quantum Materials Database (OQMD) \cite{saal2013materials,kirklin2015open} and the Materials Project \cite{jain2013commentary}, has further extended the applicability of ML-based modeling to a broad range of materials systems. However, interfacial structure searches such as MHM inherently involve numerous off-equilibrium configurations encountered during the exploration process. These configurations often correspond to transition states or high-energy structures that are underrepresented in uMLIP training data. As a result, uMLIPs may predict an incorrect energy landscape under such conditions, leading to fictitious low-energy states and unreliable energy rankings. Consequently, system-specific MLIPs remain indispensable for reliably capturing the structural and energetic complexity of interfacial systems.

In this work, we present an efficient and accurate hybrid approch for atomic-scale interface structure prediction by integrating the MHM \cite{chou2025predicting} with the Allegro MLIP \cite{musaelian2023learning}. The proposed MHM–Allegro approach is demonstrated on SrTiO$_3$ tilt grain boundaries, where candidate configurations generated by MHM are refined using DFT to evaluate interfacial energetics. We benchmark the performance of the approach against established methods and show that it reliably identifies ground-state configurations without requiring iterative MLIP retraining. The predictive capability of the method is further validated by comparison with reported experimental observations of SrTiO$_3$ $\Sigma5$(310)[001] grain boundaries, where the predicted atomic structures exhibit strong agreement with experimental measurements. Overall, this study establishes the MHM–Allegro approach as a robust and generalizable methodology for predictive modeling of complex material interfaces.

\section{\label{sec:meth}METHODS AND COMPUTATIONAL DETAILS}

The fundamental principles of the MHM for interface prediction are detailed in Reference \cite{chou2025predicting}. Section~\ref{sec:MA} presents the workflow of the current MHM implementation, including the construction of grain boundary supercells, the generation of diverse initial configurations through rigid-body translations and compositional variations, and the post-processing procedures. The computational parameters employed in the DFT relaxations, together with the formulation of interfacial energy, are described in Section~\ref{sec:MB}. Finally, Section~\ref{sec:MC} details the construction of the defect dataset for the MLIP and the subsequent evaluation of the Allegro potential.

\subsection{\label{sec:MA} Interfacial Structure Search Workflow}

\begin{figure}
\includegraphics[width=1.0\textwidth]{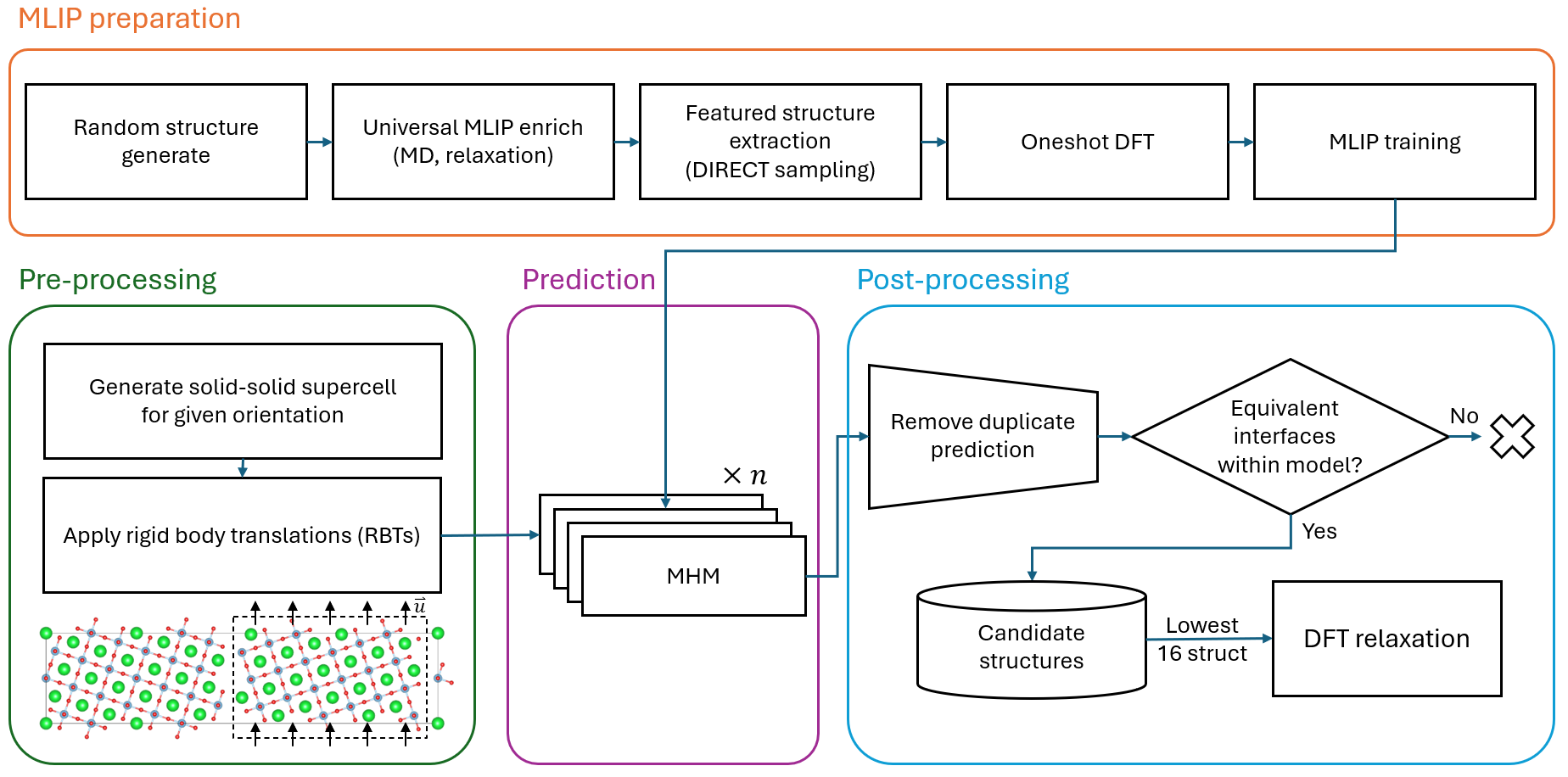}%
\caption{Workflow of MLIP preparation and MHM-based interface structure search.}
\label{fig:one}
\end{figure}

The interfacial structure-search workflow employed in this study is schematically illustrated in Figure~\ref{fig:one}. For all interface prediction methods, selecting an appropriate initial configuration is critical to ensure efficient exploration of the configurational space. In this work, we focus on tilt grain boundaries, which are modeled by constructing two-dimensionally periodic slabs representing grains of finite thickness. These slabs are then joined to form a planar grain boundary supercell. The initial tilt angles for the systems examined in this study are $70.5^\circ$ for $\Sigma3$(112)[110] and $36.9^\circ$ for $\Sigma5$(310)[001], respectively. The lattice constants obtained from DFT-optimized tetragonal SrTiO$_3$ are used to define the cell vectors of the initial configurations. Because the MHM performs structural reconstruction primarily near the interface region, distinct starting configurations for a given boundary are generated by applying systematic rigid-body translations (RBTs) between the two grains. For non-stoichiometric boundaries, additional SrO or TiO$_2$ units (\( N_{\text{SrO}} \) or \( N_{\text{TiO}_2} \)) are randomly introduced into the interfacial vacuum region of the stoichiometric model to introduce compositional variations. Apart from specifying the tilt angle and constructing the corresponding slab models, no additional heuristic design or interface-specific structural motifs are imposed. Consequently, all initial configurations are free from manual tuning or prior assumptions regarding the interfacial structure. Multiple MHM simulations are then executed in parallel, each initialized from a different configuration. Upon completion, duplicate structures were removed, and only unique GB configurations were retained. An additional filtering step was performed to confirm that each supercell contained two symmetrically equivalent interfaces. This constraint simplifies subsequent analysis, as the interfacial energy of a single interface can be obtained directly by dividing the excess energy relative to the reference state by two. This equivalence was verified through a two-step procedure. First, the supercell was required to exhibit a space group symmetry higher than P1, indicating potential equivalence between the two interfaces. Second, two planar cuts perpendicular to the grain boundary plane were introduced at an arbitrary position and at another position shifted by 50 percent of the supercell length along the interface-normal direction. Owing to the periodic boundary conditions, these two cuts effectively bisect the original supercell into two substructures. These substructures were compared using the \texttt{StructureMatcher} function implemented in the \texttt{Pymatgen} library~\cite{ong2013python}. If the two interfaces in the original supercell were equivalent, the substructures differed only by a symmetry operation (e.g., translation or rotation), and \texttt{StructureMatcher} identified them as structurally identical.

The MHM is coupled with \texttt{LAMMPS}~\cite{thompson2022lammps}, which is used to compute the energies, forces, and stresses required during the search process. These quantities are passed back to MHM, which performs molecular dynamics (MD) escape trials followed by local relaxation. For this study, we compiled a version of \texttt{LAMMPS} that supports the Allegro potential~\cite{musaelian2023learning}, enabling the MHM to be driven entirely by machine learning interatomic potentials. During MHM searches, the cell vectors lying in the grain boundary plane are fixed, while the lattice vector normal to the boundary is left unconstrained to allow relaxation in that direction. Structural relaxations within the MHM are considered converged when all atomic force components fall below 0.005~eV/\AA. The order-parameter-based biasing method~\cite{chou2025predicting} is employed, using a cutoff radius of 8~\AA~to define the local atomic environment.

\subsection{\label{sec:MB}DFT relaxation and interfacial energy}

After completing all MHM simulations and subsequent structural filtering, the sixteen lowest-energy candidate structures (evaluated using the surrogate PES models) were selected for further relaxation using DFT. The surrogate PES model employed in this study was either the Buckingham potential \cite{chou2025predicting} or the Allegro potential. The DFT calculations were implemented in the Vienna Ab Initio Simulation Package (VASP)~\cite{kresse1993ab,kresse1994ab,kresse1996efficiency,kresse1996efficient}. The Local Density Approximation (LDA) and the Projector Augmented Wave (PAW) method were used in all DFT calculations. The \texttt{Sr\_sv} and \texttt{Ti\_sv} pseudopotentials were employed to explicitly include the semicore states, namely 4s$^2$4p$^6$5s$^2$ for Sr and 3s$^2$3p$^6$3d$^2$4s$^2$ for Ti, respectively. The LDA was chosen because generalized gradient approximation (GGA) functionals are known to systematically underestimate surface energies, whereas the overbinding tendency of LDA partially cancels these errors and has been shown to yield more accurate interfacial energies~\cite{benedek2008interatomic}. A plane-wave energy cutoff of 520~eV was adopted for all calculations. The Brillouin zone was sampled using a $\Gamma$-centered \textit{k}-point mesh with a reciprocal-space resolution of 0.5~\AA$^{-1}$. Structural relaxations were performed using the conjugate-gradient algorithm, with convergence achieved when all atomic force components fell below 0.01~eV/\AA. During relaxation, the in-plane lattice parameters (parallel to the grain-boundary plane) were fixed, while the out-of-plane lattice vector and all atomic positions were allowed to relax.

The interfacial energy at the DFT level is computed using the following expression~\cite{chua2010genetic,schusteritsch2014predicting,zhao2014interface,chou2025predicting}:
\begin{equation}
\sigma = \frac{1}{2A} \left( G_{\text{GB}} - N_{\text{SrO}} \mu_{\text{SrO}} - N_{\text{TiO}_2} \mu_{\text{TiO}_2} \right),
\label{eq:one}
\end{equation}
where \( A \) is the grain boundary cross-sectional area, and \( G_{\text{GB}} \) denotes the total energy of the relaxed grain boundary structure obtained from DFT. The terms \( N_{\text{SrO}} \) and \( N_{\text{TiO}_2} \) are the numbers of SrO and TiO\(_2\) units, respectively, in the supercell. The chemical potentials of the binary components are defined as:
\begin{eqnarray}
\mu_{\text{TiO}_2} = g_{\text{TiO}_2}^{0} + \lambda \Delta G_{f.SrTiO_3}^{0}, \label{eq:two} \\
\mu_{\text{SrO}} = g_{\text{SrO}}^{0} + (1 - \lambda) \Delta G_{f.SrTiO_3}^{0}, \label{eq:three}
\end{eqnarray}
where \( 0 \leq \lambda \leq 1 \), and \( g_{\text{SrO}}^{0} \) and \( g_{\text{TiO}_2}^{0} \) are the DFT total energies of the ground-state rock salt SrO and rutile TiO\(_2\) phases, respectively. The formation energy of SrTiO\(_3\) is given by:
\begin{equation}
\Delta G_{f.SrTiO_3}^{0} = g_{\text{SrTiO}_3}^{0} - g_{\text{SrO}}^{0} - g_{\text{TiO}_2}^{0},
\end{equation}
where \( g_{\text{SrTiO}_3}^{0} \) is the DFT total energy of bulk tetragonal SrTiO\(_3\). The calculated value of $\Delta G_{f.SrTiO_3}^{0}$ is -1.40~eV.

Substituting Eqs.2 and 3 into Eq.1, the interfacial energy can be rewritten as:
\begin{widetext}
\begin{equation}
\sigma = \frac{1}{2A} \left( G_{\text{GB}} - N_{\text{SrO}} g_{\text{SrO}}^{0} - N_{\text{TiO}_2} g_{\text{TiO}_2}^{0} - N_{\text{SrO}} \Delta G_{f.SrTiO_3}^{0} + X N_{\text{SrO}} - X N_{\text{TiO}_2} \right),
\label{eq:four}
\end{equation}
\end{widetext}
where \( X = \mu_{\text{TiO}_2} - g_{\text{TiO}_2}^{0} = \lambda \Delta G_{f.SrTiO_3}^{0} \) serves as a tunable chemical potential variable. It should be noted that the bulk region is not modified across simulations; rather, the variation in chemical potential is represented through the reference parameter \(X\). This formulation enables comparison of stoichiometric and non-stoichiometric grain boundaries across a range of chemical potentials. To systematically classify the stoichiometry of each boundary model, we define a non-stoichiometry descriptor \( \Gamma \) as:
\begin{equation}
\Gamma = \frac{N_{\text{TiO}_2} - N_{\text{SrO}}}{2\tilde{N}_{\text{SrTiO}_3}},
\end{equation}
where \( \tilde{N}_{\text{SrTiO}_3} \) represents the number of SrTiO\(_3\) formula units per atomic layer parallel to the grain boundary plane. In this scheme, stoichiometric boundaries are characterized by \( \Gamma = 0 \).

\subsection{\label{sec:MC} Dataset preparation and MLIP training}

\begin{figure}
\includegraphics[width=1.00\textwidth]{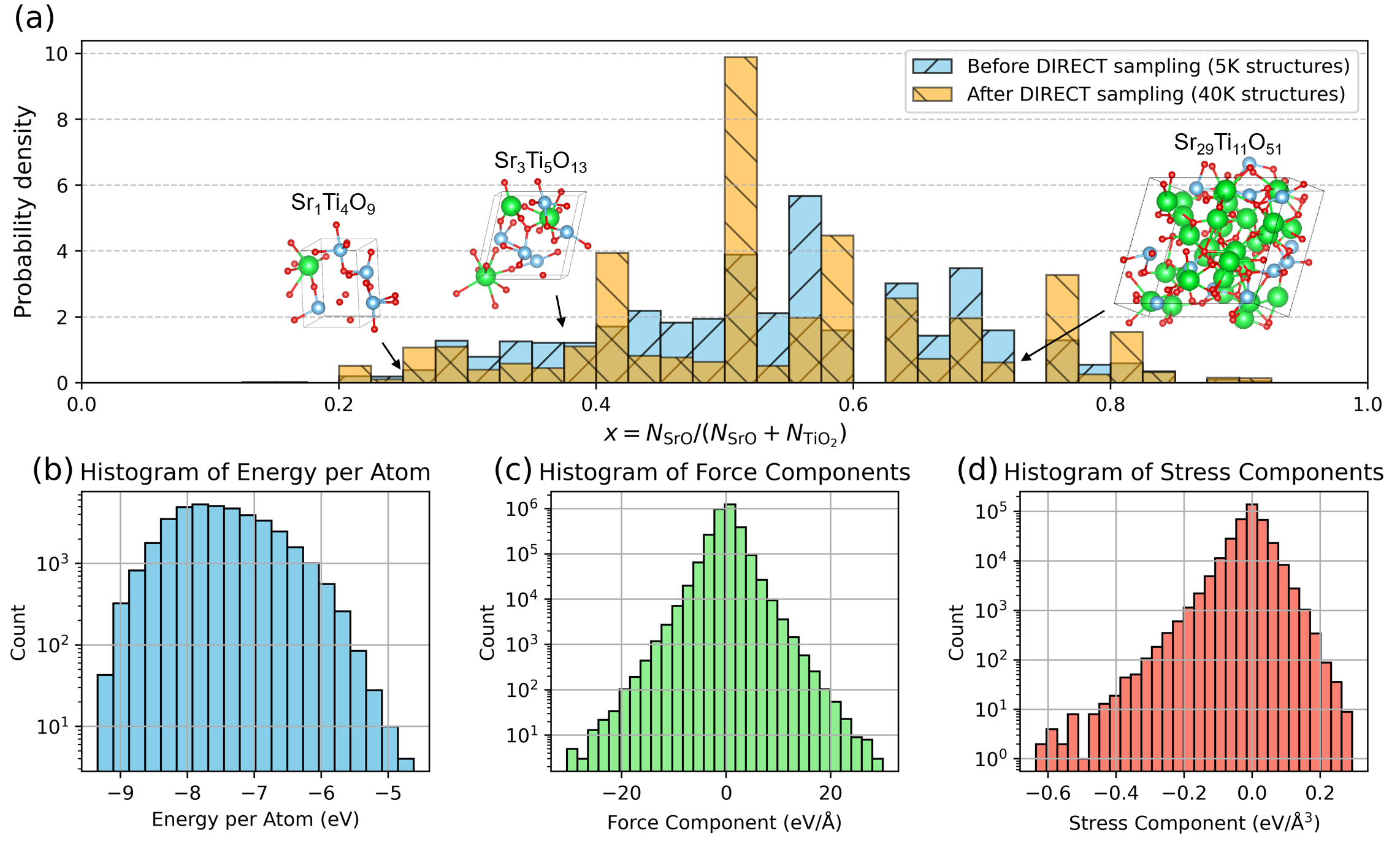}%
\caption{The dataset for MLIP. (a) Initial randomly generated structures and final selected configurations; and (b)–(d) the energy, force, and stress components in the final DFT dataset after DIRECT sampling}
\label{fig:two}
\end{figure}

The MLIP preparation workflow is illustrated in the top panel of Figure~\ref{fig:one}. Instead of tailoring the training data to specific atomic configurations, we constructed a randomized dataset to encourage the MLIP to learn general interatomic interactions among Sr, Ti, and O atoms, enabling its application to a wide range of atomistic simulations, including the interface prediction task considered in this study. The grain boundary systems examined here span the compositional range \( x\,\text{SrO} + (1 - x)\,\text{TiO}_2 \), with denser sampling near the stoichiometric composition (\( x = 0.5 \)). We began by randomly generating 5000 chemically diverse initial configurations across this range, each containing fewer than 100 atoms per unit cell. The generation process followed geometric constraints such as interatomic distance limits, and each configuration was assigned a physically reasonable cell volume proportional to the number of SrO and TiO\(_2\) units, calculated as  \( V_{\text{cell}} = N_{\text{SrO}}V_{\text{SrO}} + N_{\text{TiO}_2}V_{\text{TiO}_2} \), where \( V_{\text{SrO}} \) and \( V_{\text{TiO}_2} \) are their respective approximate atomic volumes evaluated from their ground-state structures. These randomly generated configurations are shown as blue bins in Figure~\ref{fig:two}(a). After generation, we employed the uMLIP \textbf{M3GNet}~\cite{chen2022universal,kaplan2025foundational}—specifically the model \texttt{M3GNet-MatPES-PBE-v2025.1-PES}, available from the MatGL repository~\cite{ko2025materials}—to perform MD simulations and subsequent structural relaxations for each configuration. The objective of this stage was to enrich configurational diversity while maintaining physical plausibility. The uMLIP-based MD simulations were conducted in the canonical (NVT) ensemble with a time step of 1.0~fs for a total of 500 steps at 700~K. Following the MD runs, the same M3GNet potential was employed to relax the final configurations from each trajectory using the FIRE optimizer~\cite{bitzek2006structural}, with convergence defined by a maximum atomic force below 0.05~eV/\AA~or 200 relaxation steps. Each initial configuration yielded approximately 700 derived configurations from the MD and relaxation stages, resulting in a cumulative dataset of about 3.5~million configurations. To extract a representative subset, we applied DIRECT sampling~\cite{qi2024robust}. DIRECT sampling is a dimensionality-reduction and stratified clustering approach designed to efficiently select a diverse and representative subset of structures from a large configuration space. To address memory constraints encountered during clustering in the DIRECT sampling, a two-step sampling strategy was implemented. In the first step, the 5000 trajectories were divided into 50 batches, and DIRECT sampling was performed separately on each batch, reducing approximately 70,000 configurations to 20,000 configurations per batch. The resulting subsets were then combined into a reduced pool of about 1~million configurations. In the second step, DIRECT sampling was again applied to this pooled dataset to obtain a final set of 40,000 diverse configurations. The average coverage scores of the first-round sampling across all batches consistently exceeded 0.99, while the second-round sampling achieved an average score of 0.966. Finally, these 40,000 selected structures were calculated using oneshot DFT calculations with the same computational settings described in Section~\ref{sec:MB}, forming the final dataset at the DFT level. These selected configurations are shown as yellow bins in Figure~\ref{fig:two}(a).

Figure~\ref{fig:two}b--d illustrate the distributions of energy, force, and stress within the dataset. These distributions are well spread, suggesting a diverse sampling of the configuration space that is expected to provide essential information for the MLIP to accurately capture the interactions among Sr, Ti, and O atoms. In this work, we employ the Allegro~\cite{musaelian2023learning} potential (version 0.5.0) as the MLIP model for interface structure search. The dataset is divided into training, validation, and test sets with a ratio of 0.8:0.1:0.1. The Allegro model architecture comprises two layers with 32 features for both even and odd irreducible representations and $\ell_{\text{max}} = 3$. The Allegro multilayer perceptron (MLP) includes three hidden layers, each with 128 units, and employs the SiLU activation function. The scalar embedding MLP consists of two hidden layers with 32 features and also uses the SiLU activation. A radial cutoff of 6.0~\AA{} is applied, along with eight non-trainable Bessel functions for radial basis encoding. The polynomial envelope function uses an exponent of $p = 6$. The model is trained with a composite loss function comprising energy, force, and stress terms, each weighted equally (i.e., weight = 1). Energy contributions are evaluated as per-atom mean squared error (MSE). Training is conducted using the Adam optimizer~\cite{kingma2014adam} in PyTorch~\cite{paszke2019pytorch} with default $\beta$ parameters. The learning rate is initialized at 0.002, and the batch size is set to 5. A learning rate scheduler reduces the learning rate upon validation-loss plateaus, with a patience of 20 epochs and a decay factor of 0.5. An exponential moving average with a smoothing factor of 0.99 is employed to evaluate performance on the validation set and define the final model. Training is terminated upon satisfaction of any of the following criteria: (a) a maximum training time of 30~days, (b) a maximum of 10,000~epochs, (c) no improvement in validation loss for 200 consecutive epochs, or (d) the learning rate falls below $10^{-6}$. All training operations are performed using \texttt{float32} precision. Model training and all subsequent simulations in \texttt{LAMMPS} are executed on NVIDIA H100 GPUs.

\begin{table}[b]%The best place to locate the table environment is directly after its first reference in text
\caption{\label{tab:table2}%
The mean absolute errors (MAEs) for total energy, energy per atom, force, and stress for Allegro potential training, validation and testing
}
\begin{ruledtabular}
\begin{tabular}{ccccc}
\textrm{}&
\textrm{Energy MAE (eV)}&
\textrm{E per atom MAE (eV)}&
\textrm{Force MAE (eV/Ang)}&
\textrm{Stress MAE (eV/Ang$^3$)}\\
\colrule
Training   & 0.109 & 0.0046 & 0.089 & 0.0019 \\
Validation & 0.106 & 0.0046 & 0.108 & 0.0019 \\
Testing    & 0.111 & 0.0047 & 0.107 & 0.0019 \\
\end{tabular}
\end{ruledtabular}
\end{table}

Table~\ref{tab:table2} summarizes the training, validation and testing performance of the Allegro model. The mean absolute errors (MAEs) for total energy, energy per atom, and stress all fall within high-accuracy thresholds, indicating robust model learning of these quantities. In contrast, the MAE for atomic forces is comparatively higher and exceeds the convergence criterion typically used for structural relaxation. This larger force error can be attributed to the significant presence of high-force configurations in the training dataset, particularly those sampled from MD trajectories, as illustrated in Figure~\ref{fig:two}(c). The force components in the dataset span a broad range, approximately from $-30$ to $30$~eV/\AA{}, which increases the variance and inherently raises the achievable MAE. Considering this wide distribution, a force MAE of approximately 0.1~eV/\AA{} is regarded as reasonable. Nevertheless, our interface prediction workflow includes a final DFT structural relaxation step following the MHM search. Therefore, the model’s ability to yield fully relaxed geometries is not strictly required at this stage, and a moderately higher force MAE remains acceptable within the context of our methodology.

\section{\label{sec:result}RESULTS AND DISCUSSIONS}

\subsection{\label{sec:r112} $\Sigma3$(112)[110]: Benchmarking Computational Interface Prediction Methods}

\begin{figure}
\includegraphics[width=1.00\textwidth]{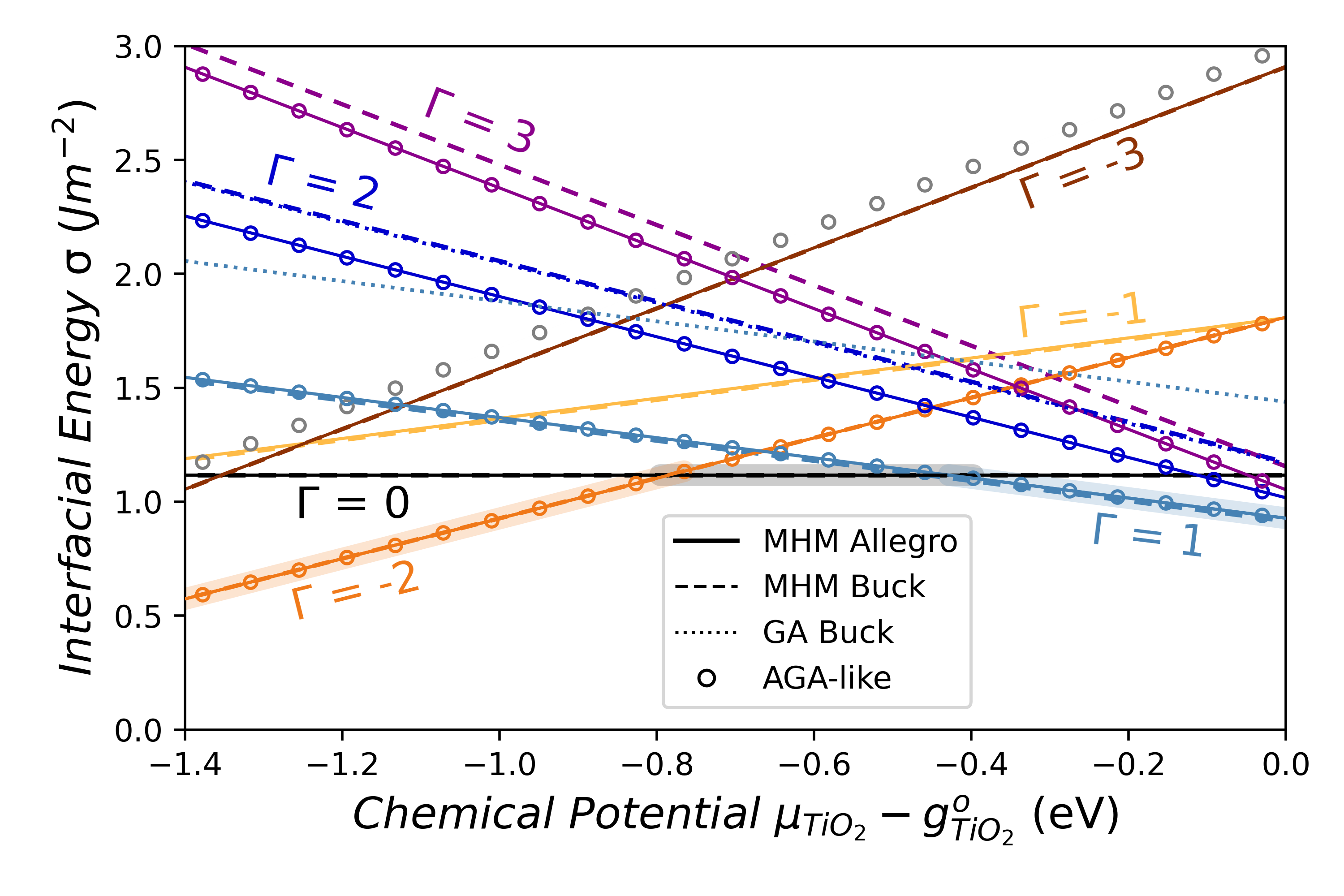}%
\caption{Interfacial energies of the $\Sigma$3(112)[110] SrTiO$_3$ grain boundaries as a function of the chemical potential of the TiO$_2$ component. The first term in the legend denotes the structure search algorithm, and the second term specifies the surrogate PES model employed during the search. All interfacial energies were evaluated at the DFT level. Wherever available, reference structures were recalculated using the same DFT settings as employed in this work to eliminate numerical inconsistencies. Because explicit atomic structures from the AGA study were not provided, we identified MHM-predicted structures that are as consistent as possible with the reported AGA results, based on the evidence presented in the Supporting Information. These structures were therefore assigned as AGA reference structures, and their interfacial energies were recalculated accordingly. In some cases, the lowest-energy structure identified by MHM coincides with the AGA reference structure, as indicated by overlapping markers and the MHM line style. The lowest-energy grain boundary at a given chemical potential is highlighted by a color-matched halo.}

\label{fig:result2}
\end{figure}

The $\Sigma3$(112)[110] GB system has been extensively investigated using various simulation approaches. In this study, we predict the structures of SrTiO$_3$ $\Sigma3$(112)[110] GBs across a range of stoichiometries ($\Gamma = -3$~to~3$)$ to assess their thermodynamic stability. Figure~\ref{fig:result2} shows the DFT-calculated interfacial energies as a function of chemical potential, evaluated using Equation~\ref{eq:four}. Only the lowest-energy configurations predicted by each method are included. At the high end of the chemical potential range (right side of the figure), corresponding to TiO$_2$-rich conditions, TiO$_2$-rich grain boundaries ($\Gamma > 0$) are thermodynamically more stable than their SrO-rich counterparts ($\Gamma < 0$). Conversely, under SrO-rich conditions (more negative chemical potentials, left side of the figure), the stability of SrO-rich boundaries increases. Depending on the chemical potential, the most stable interface structures within the SrTiO$_3$ $\Sigma3$(112)[110] system correspond to $\Gamma = -2$, $0$, and $1$ GBs.

The $\Sigma3$(112)[110] system is commonly used as a benchmark to evaluate the performance of various interface structure prediction methods~\cite{chua2010genetic,zhao2014interface,chou2025predicting}, particularly based on whether a method can identify a new lowest-energy structure or correctly reproduce a previously reported one. In the absence of an experimental ground truth, the lowest-energy configuration predicted by all methods is typically regarded as the theoretical ground truth for the given system. However, determining the lowest-energy configuration solely by comparing interfacial energies across different prediction methods is nontrivial and prone to numerical inconsistencies~\cite{chou2025predicting}. Variations in DFT software packages and computational settings can introduce discrepancies not only in the GB total energies but also in the reference states. Therefore, whenever possible, it is essential to recalculate interfacial energies using consistent computational settings. For the $\Sigma3$(112)[110] system, we recalculated the interfacial energies of the GA structures with $\Gamma = 1$ and $\Gamma = 2$ using the DFT relaxation settings described in the Methods section. The corresponding GA configurations were provided directly by the authors~\cite{chua2010genetic}. In contrast, the AGA study~\cite{zhao2014interface} does not provide explicit atomic structure files. We therefore compared our MHM-predicted configurations with the available information reported in the AGA publication. Specifically, structural overlay analyses were performed between our predicted structures and the grain-boundary images presented in the AGA study, in conjunction with comparisons of the reported interfacial energy values. Based on this combined evidence, we identified MHM-predicted structures that are as consistent as possible with the AGA results and assigned them as AGA reference structures. Details of the overlay analysis are provided in the Supplementary Information (SI)~\cite{SI}. These AGA reference structures, denoted as AGA-like structures, are indicated by additional hollow-circle symbols in Figure~\ref{fig:result2}. In some cases, the lowest-energy structure identified by MHM coincides with the AGA reference structure, as indicated by overlapping markers and the MHM line style. It should be noted that the AGA study reports grain-boundary images only for stoichiometries $\Gamma = 1$ to $3$ and $-2$ to $-3$; accordingly, our identification of AGA-consistent structures is limited to these stoichiometries. For transparency and reproducibility, all MHM-predicted grain-boundary structures discussed in this work are provided in the SI~\cite{SI}. The $\Sigma3$(112)[110] $\Gamma = 0$ grain-boundary configuration is well established and consistently reported across different methods~\cite{chua2010genetic,zhao2014interface,chou2025predicting}. We therefore assume that all approaches identify equivalent structures for this stoichiometry, and only the MHM $\Gamma = 0$ configuration is shown in Figure~\ref{fig:result2}.

% MHM with other 
As shown in Figure~\ref{fig:result2}, the MHM approach combined with the Allegro potential demonstrates superior performance compared to other methods, consistently identifying the most stable interface structures across all stoichiometries investigated. For each GBs previously identified by the GA method, MHM with the Allegro potential was able to find configurations with lower interfacial energies. Both the AGA and MHM with the Allegro potential identified the most stable interfaces for $\Gamma = 1$ to $3$ and $\Gamma = -2$. However, for $\Gamma = -3$, more stable structures were obtained using either MHM with the Allegro potential or MHM with the Buckingham potential. The line formed by the gray circle markers in Figure~\ref{fig:result2} indicates the $\Gamma = -3$ structure identified by AGA, which lies 0.09~J/m$^{2}$ higher in energy than the MHM-predicted structures. Notably, the Allegro potential used in this study was trained solely on randomly generated structures, without iterative refinement or inclusion of interface-specific data. Even with this general-purpose training, the MHM–Allegro approach consistently identified the most stable configurations across all stoichiometries. Furthermore, the lowest-energy structures predicted by the Allegro potential are consistently preserved as the lowest-energy configurations after subsequent DFT relaxation. Together, these results demonstrate both the strong extrapolation capability of the potential and its high accuracy in capturing the relative energetics of complex grain-boundary configurations. In our previous work~\cite{chou2025predicting}, we also investigated the $\Sigma3$(111)[110] GB system. Here, we re-predicted the $\Sigma3$(111)[110] GBs using MHM with the same Allegro potential, and the results are provided in the SI~\cite{SI}. Overall, MHM combined with the Allegro potential reliably identifies the lowest-energy interface structures across both the $\Sigma3$(112)[110] and $\Sigma3$(111)[110] systems, demonstrating its stability and computational efficiency in exploring complex grain boundary configurations.

% MHM with different potential
The most significant improvements are observed for $\Gamma > 0$ when the potential is switched from Buckingham to Allegro within the MHM. In contrast, for $\Gamma = -1$ to $-3$, MHM with both potentials identifies the same low-energy configurations. This result can be rationalized by considering that negative $\Gamma$ values correspond to SrO-rich grain boundaries, where the most stable SrO bulk phase—rock-salt—is relatively simple and can be accurately described by a classical Buckingham potential. In comparison, positive $\Gamma$ values are associated with TiO$_2$-rich interfaces, where the complex PES of TiO$_2$ poses a greater challenge for Buckingham potentials. In these cases, the Allegro potential captures the underlying physics more precisely, leading to notable improvements in interface prediction. Notably, in our previous work~\cite{chou2025predicting}, the AGA $\Gamma = -3$ grain boundary could not be identified using MHM with the Buckingham potential. In contrast, the same low-energy configuration was successfully recovered using MHM combined with the Allegro potential in the present study, highlighting that the prediction of SrO-rich GBs is also improved. Finally, we tested the direct integration of the M3GNet uMLIP~\cite{chen2022universal,kaplan2025foundational} into the MHM for SrTiO$_3$ benchmark systems; however, this approach failed to reproduce the lowest-energy GB structures obtained with the Allegro potential for the $\Sigma3$(112)[110] system, underscoring the necessity of developing system-specific MLIPs for accurate interface prediction. 

\begin{figure}
\includegraphics[width=1.00\textwidth]{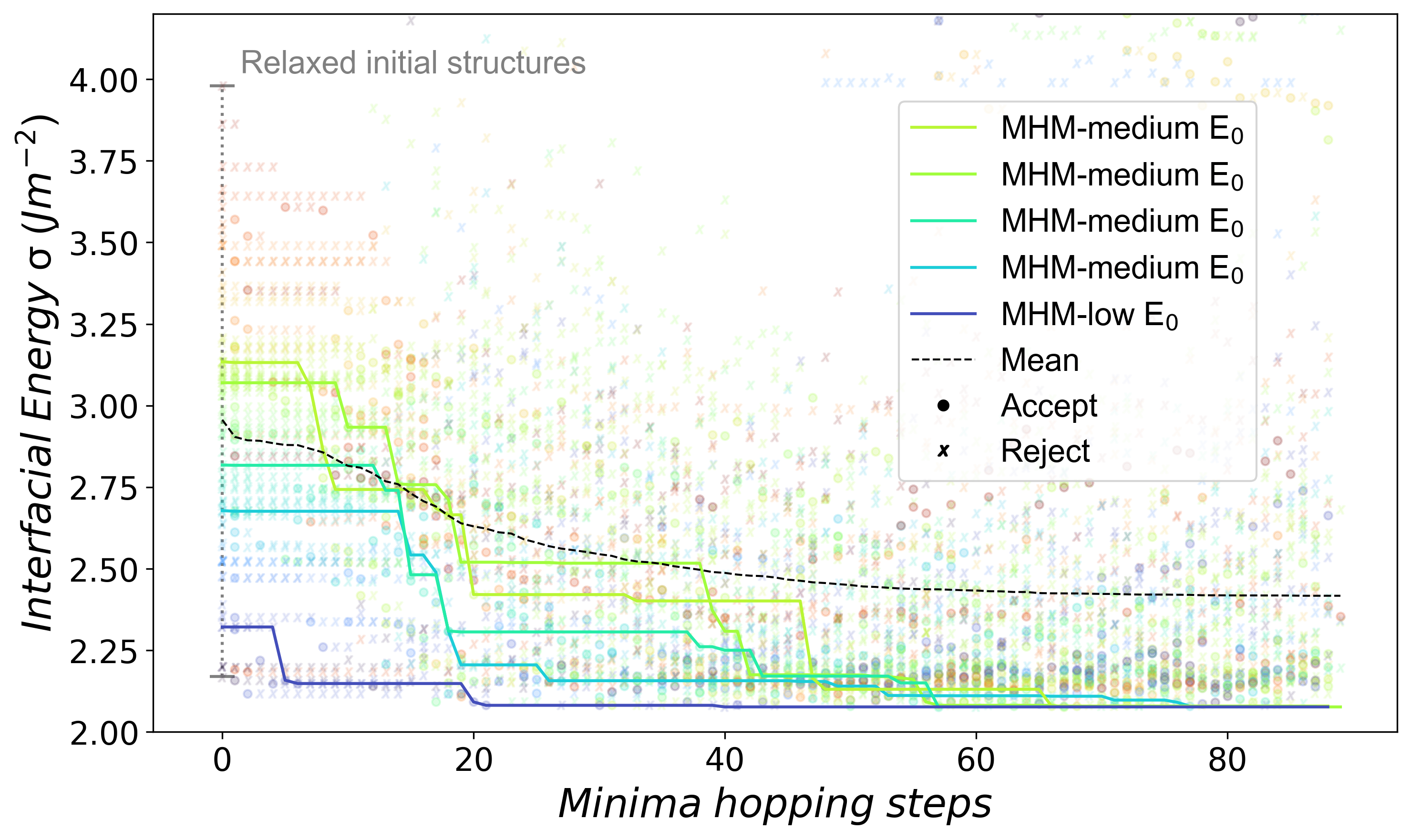}%
\caption{Representative example of the interface-structure search workflow. Ninety independent MHM runs were performed in parallel for the SrTiO$_3$ $\Sigma$3(112)[110] $\Gamma = 3$ grain boundaries. Interfacial energies were evaluated at $\mu_{\mathrm{TiO_2}} - g_{\mathrm{TiO_2}}^{0} = -0.7$ eV using a trained Allegro potential. Five representative MHM trajectories are highlighted and ultimately converge to the lowest-energy configuration for $\Gamma = 3$; four out of five originate from medium-energy starting points.}
\label{fig:result3}
\end{figure}

To evaluate the effectiveness of MHM as a structure-search algorithm, the search process for the $\Gamma = 3$ interface is illustrated in Figure~\ref{fig:result3} as a representative example. All interfacial energies were evaluated using the Allegro potential. Each MHM run explores local minima and determines whether to accept a newly found structure as the next starting point (dot markers) or to reject it and resume from the previous minimum (cross markers). A total of 90 independent MHM searches were executed in parallel, each initialized with different RBTs and randomly inserted atoms for non-stoichiometric grain boundaries. Each search began with a relaxation of the initial configuration, with the resulting energy recorded as step zero. The gray band at step zero in Figure~\ref{fig:result3} represents the energy range of all initially relaxed structures. Although some of these configurations exhibit relatively low interfacial energies, they do not correspond to the true minima subsequently identified by MHM. Five representative MHM trajectories are highlighted, tracing the lowest energy reached at each hopping step and ultimately converging to the lowest-energy configuration for $\Gamma = 3$. The dashed line in Figure~\ref{fig:result3} represents the mean of the lowest energy reached up to each hopping step among all independent MHM runs. Figure~\ref{fig:result3} clearly demonstrates that MHM reliably identifies lowest-energy interfaces beyond random structure-search approaches. Notably, four of these highlighted trajectories originated from medium-energy starting points, underscoring the robustness of MHM in locating the global minimum even from suboptimal initial conditions. In this particular $\Gamma = 3$ case, more MHM searches were performed than strictly necessary to locate the lowest-energy configuration. 

\subsection{\label{sec:r310}SrTiO$_3$ $\Sigma5$(310)[001]: Case Study of MHM Predictions and Experimental Measurements}

\begin{figure}
\includegraphics[width=1.00\textwidth]{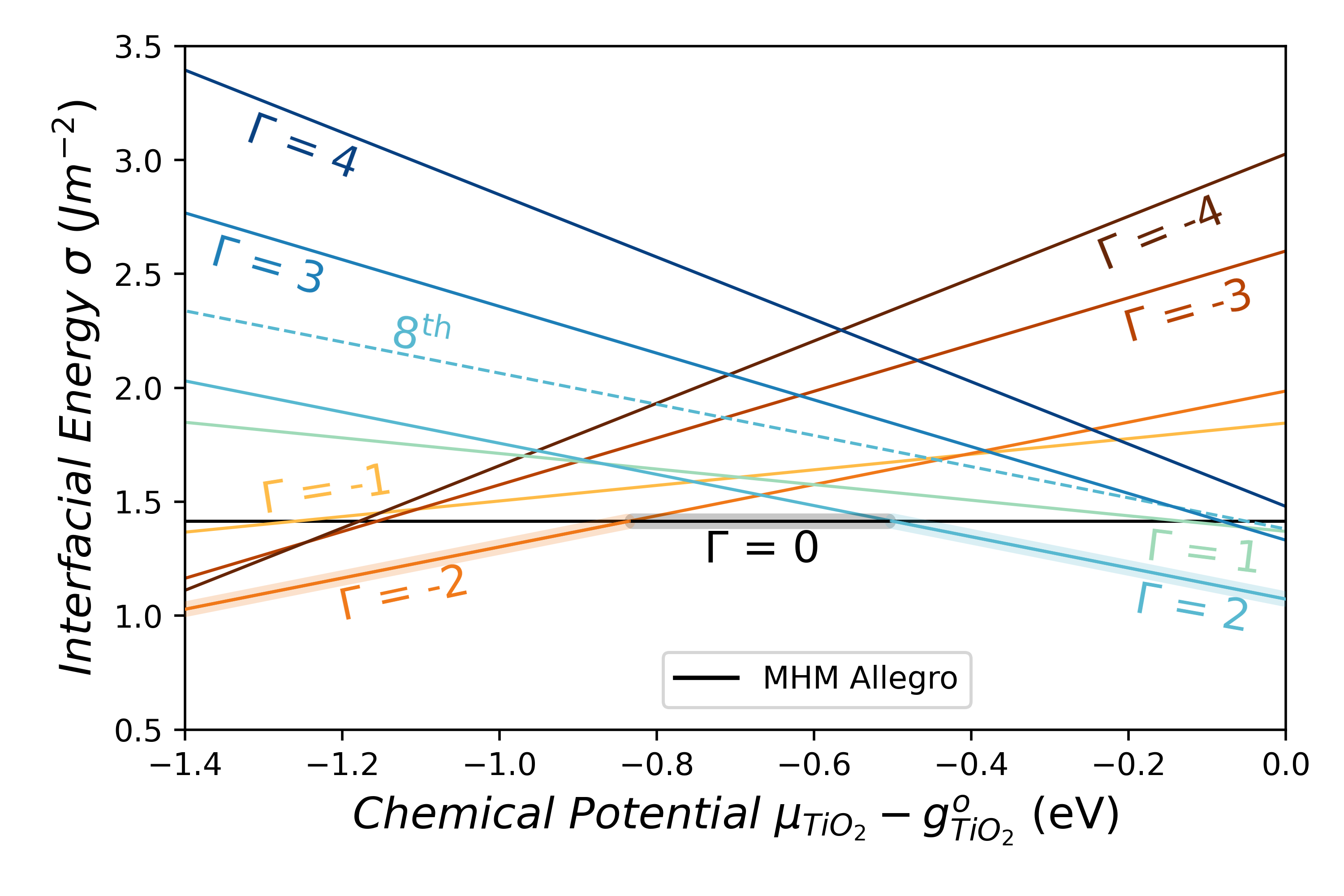}%
\caption{Interfacial energies of the $\Sigma$5(310)[001] SrTiO$_3$ grain boundaries as a function of the chemical potential of the TiO$_2$ component. All interfacial energies are calculated at the DFT level. The lowest-energy grain boundary at a given chemical potential is highlighted by a color-matched halo.}
\label{fig:310}
\end{figure}

To evaluate the predictive capability of our MHM–Allegro approach, we applied it to the SrTiO$_3$ $\Sigma5$(310)[001] grain boundary, one of the most extensively studied oxide interfaces~\cite{yang2021determination,liao2021nanoscale,ravikumar1993atomic,ravikumar2000atomic,dravid2000atomic,imaeda2008atomic,lee2003temperature}. Experimental studies have reported the existence of multiple distinct stoichiometric variants of this grain boundary. Based on our analysis of the atomic structures presented in these reports, the corresponding non-stoichiometry descriptor spans approximately from $\Gamma = 0$ to $\Gamma = 4$ (this descriptor captures only the imbalance between TiO$_2$ and SrO units and therefore reflects cation non-stoichiometry; variations in oxygen content (e.g., oxygen vacancies) are not captured). By contrast, theoretical investigations have primarily focused on the stoichiometric case ($\Gamma = 0$), largely because it has been assumed to represent the most thermodynamically stable configuration. In this study, we systematically explore a significantly wider stoichiometric space, covering $\Gamma = -4$ to $\Gamma = 4$, and identify the lowest-energy GB configuration within each stoichiometry class using the same MHM–Allegro approach shown in previous result section. Figure~\ref{fig:310} demonstrates that the stoichiometric GB is not the lowest-energy configuration across the entire range of chemical potentials. Instead, SrO-rich and TiO$_2$-rich GBs emerge as energetically favored under their respective chemical potential limits. The predicted $\Gamma = 0$ GB structure reproduces the configuration reported by Imaeda \textit{et al.}~\cite{imaeda2008atomic}. When recalculated using the same GGA pseudopotentials, our interfacial energy (0.98~J\,m$^{-2}$) closely matches their reported value (1.02~J\,m$^{-2}$), confirming that both studies describe the same atomic configuration. The cation positions in our model also show excellent consistency with their atomic-scale images. All predicted GB structures are provided in the SI~\cite{SI}.

Although Imaeda \textit{et al.} established that the stoichiometric configuration agrees well with HAADF–STEM observations, the TiO$_2$-rich GB reported by Yang \textit{et al.}~\cite{yang2021determination} remains structurally unresolved. This lack of an atomic-scale structural model limits the interpretation of their observed space-charge layer. To address this, we compare our predicted TiO$_2$-rich GB structures with their EELS measurements to identify which configurations are most consistent with the experimental observation. Before presenting this comparison, it is important to clarify several fundamental differences between our modeling assumptions and the experimental conditions. First, Yang \textit{et al.} reported partial reduction of Ti in the GB core, corresponding to an average oxidation state of approximately +3.75 and indirectly suggesting the presence of oxygen deficiency. In contrast, all grain-boundary models considered in this study assume overall charge neutrality and explicitly exclude oxygen vacancies. This assumption is adopted because many current MLIPs rely on finite local atomic environments and therefore cannot accurately capture long-range electrostatic interactions. Furthermore, the training dataset used to construct the interatomic potential consists exclusively of charge-balanced DFT configurations. A second key distinction is that the EELS measurements represent an averaged projection over a sampling depth of roughly 15~nm, meaning that the reported intensities integrate information from multiple GB segments. Yang \textit{et al.} accordingly observed mixed Sr/Ti column intensities within the GB core. In contrast, our computational models contain only a few hundred atoms and represent idealized, periodic GB structures without mesoscale variability. A further difference arises from finite-temperature effects. Although DFT-based predictions correspond to 0~K, the experimentally measured interface was formed under high-temperature heat treatment and may not strictly reflect a fully equilibrated structure, even if bicrystals are often assumed to approximate near-equilibrium grain-boundary configurations. To ensure a meaningful comparison, we therefore consider not only the lowest-energy GB structures within a given stoichiometry but also low-energy configurations that could be experimentally accessible.

\begin{figure}
\includegraphics[width=0.5\textwidth]{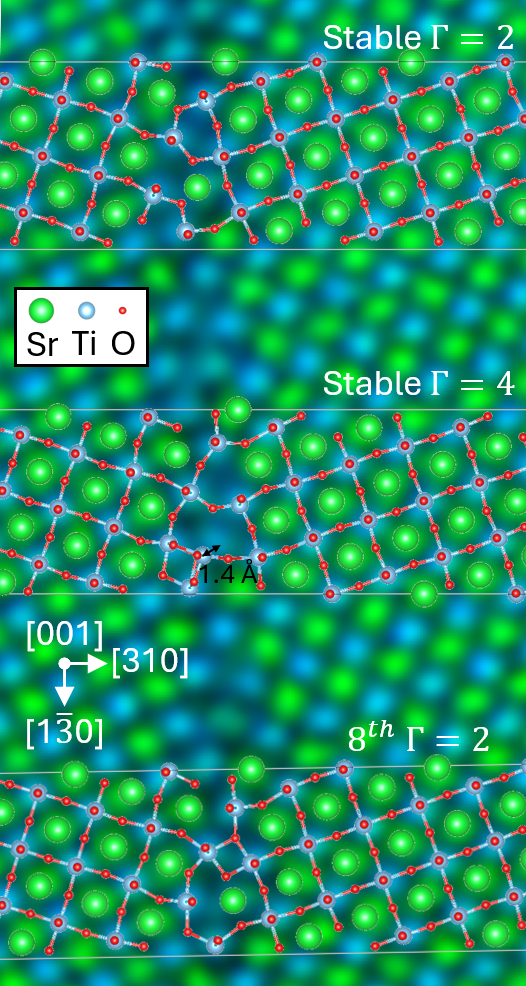}%
\caption{Overlay comparison of the most stable $\Gamma = 2$ structure, the most stable $\Gamma = 4$ structure, and the eighth-lowest-energy $\Gamma = 2$ structure predicted in this study with the EELS image reported by Yang \textit{et al.}\cite{yang2021determination}. The white horizontal lines indicate the periodic boundaries of the supercell.}
\label{fig:Temp}
\end{figure}

The experimentally observed TiO$_2$-rich interface~\cite{yang2021determination} corresponds to either $\Gamma = 2$ or $\Gamma = 4$ under our definition. Within these stoichiometries, we identify three predicted GB structures that exhibit strong agreement with the EELS projection: the lowest-energy $\Gamma = 2$ GB, the lowest-energy $\Gamma = 4$ GB, and a higher-energy $\Gamma = 2$ configuration that ranks eighth lowest energy within the $\Gamma = 2$ class. Structural overlays with the experimental projection are shown in Figure~\ref{fig:Temp}, and their interfacial energies are shown in Figure~\ref{fig:310}. The lowest-energy $\Gamma = 2$ GB is thermodynamically the most favorable among the three EELS-consistent GBs. In fact, over a finite range of chemical potentials, it also exhibits the lowest interfacial energy among all predicted $\Sigma 5$(310)[001] SrTiO$_3$ GBs .The $\Gamma = 4$ interface reproduces most projected features but shows a 1.4~Å displacement of the central Ti column relative to experiment. Notably, these experimentally consistent configurations are obtained starting from idealized and unbiased initial configurations defined solely by the tilt angle, without incorporating any experimental structural information. This highlights the predictive capability of the MHM–Allegro approach in exploring complex interfacial energy landscapes and in identifying physically relevant stable and low-energy grain-boundary structures.

The limitations of the current MHM–Allegro approach must also be acknowledged. The Allegro potential employ local atomic environments within a finite cutoff and include iterated tensor products of learned equivariant representations; consequently, they do not capture long-range electrostatic interactions with sufficient accuracy. As a result, the present Allegro potential are not yet appropriate for describing mesoscale electrostatic phenomena such as space-charge layer formation. Incorporating explicit electrostatic effects into MLIPs is an important direction for future development~\cite{kim2025universal}, particularly for oxide systems where charged defects and long-range Coulomb interactions play a central role.

\section{\label{sec:con}Conclusion}

In summary, we have developed an efficient and transferable approach for predicting atomic-scale interface structures by integrating the Minima Hopping Method with the Allegro machine-learning interatomic potential. Benchmark tests on SrTiO$_3$ tilt grain boundaries of $\Sigma3$(112)[110] demonstrate that the MHM–Allegro approach reliably identifies the most stable interface configurations across varying stoichiometries without the need for iterative MLIP refinement. The proposed strategy for constructing defect-representative training datasets—based on random structure generation, uMLIP-driven enrichment, and representative selection via DIRECT sampling—proves effective for achieving broad generality in MLIP training. Application of this approach to SrTiO$_3$ $\Sigma5$(310)[001] grain boundaries further confirms its predictive capability, with the resulting atomic structures showing excellent agreement with experimental observations. At the same time, we acknowledge that the simplified GB models and inherent limitations of current MLIPs may not fully capture the complexity of real interfaces. Collectively, these results highlight the potential of the MHM–Allegro approach to bridge the gap between \textit{ab initio} modeling and experimental characterization of material interfaces, providing a robust foundation for the systematic exploration of more complex interfacial systems.

\begin{acknowledgments}
This research was primarily supported as part of the Hydrogen in Energy and Information Sciences (HEISs), an Energy Frontier Research Center funded by the U.S. Department of Energy (DOE), Office of Science, Basic Energy Sciences (BES), under award No. DE-SC0023450. C.C. would like to acknowledge the financial support received from MOE Taiwan Scholarship at Northwestern. M.W. would like to acknowledge the funding source from National Science Foundation, Office of Advanced Cyberinfrastructure (OAC), under Award No. 2118201. This research used resources of the National Energy Research Scientific Computing Center (NERSC), a Department of Energy User Facility using NERSC award BES-ERCAP0032806. This research was supported in part through the computational resources and staff contributions provided for the Quest high performance computing facility at Northwestern University which is jointly supported by the Office of the Provost, the Office for Research, and Northwestern University Information Technology.

\end{acknowledgments}

\bibliography{references}

\end{document}